\documentclass{ltxguide}

\usepackage{color,graphicx,shortvrb}
\usepackage{amsmath,amsthm}

\chardef\bslash=`\\ 

\theoremstyle{definition}

\theoremstyle{remark}

\newcommand{\eval}[2][\right]{\relax
  \ifx#1\right\relax \left.\fi#2#1\rvert}

\let\abs=\envert

\DeclareGraphicsExtensions{.ps}

\let\package\textsf

\newlength{\gxlen}
\MakeShortVerb{\|}

\begin{document}

\title{Relaxation, Emission Modes and Limit Temperatures in Ni+Ni HIC.}
  
\author{   Armando  Barra\~n\'on  
\footnote{ Universidad Aut\'onoma Metropolitana. Unidad Azcapotzalco.
Av. San Pablo 124, Col. Reynosa-Tamaulipas, Mexico City. email: bca@correo.azc.uam.mx } ;
 J. A.  L\'opez
\footnote{Dept. of Physics, The University of Texas at El Paso. El Paso, TX, 79968  }    ;
C. Dorso.
\footnote{ Depto.de F\'{\i}sica, Universidad de Buenos Aires. Buenos Aires, Argentina}   } 

\date{November $15^{th}$  2003}

\maketitle

\abstract

A dynamical stability analysis is performed for Ni+Ni
 central collision at intermediate energies, showing that 
in chemical, thermal and dynamical equilibriums are reached 
at an early stage of system evolution. This is obtained by 
computing the relaxation times of the quadrupole momentum,
 speed of sound and electric charge density of the system. 
This way,  fragment emission modes at low and high energies 
as well as the qualitative behavior of the limit temperatures are determined.

\section{Introduction}

   In a heavy ions collision, a compound nuclear system
 is initially formed that will expand as a consequence of
 the thermal or compressive pressure. Expansion leads the
 nuclear system into the instability region where the speed of
 sound attains an imaginary value. At this instability region, 
system is disassembled into fragments and nucleons. 
Fragmentation results from an adiabatic expansion of hot
 nuclear matter produced in the collision that leads the system
 into the mechanical instability region, where the system fragments
 into clusters and nucleons due to the increase of density fluctuations \cite{1}. 
    Neutron rich Heavy ions collisions lead to transient states of nuclear
 matter with high spin asymmetry and with a great thermal and compressive
 excitation, allowing the exploration of the properties of nuclear matter
 whose neutron contain goes from pure neutron nuclear matter up to 
symmetric nuclear matter \cite{2}. Müller et. al. has shown that the liquid-gas
 phase transition in asymmetric nuclear matter is of second order instead 
of the first order phase transition observed in symmetric nuclear matter \cite{3}. 

   Several important aspects of nuclear collisions depend on the isospin, such
 as the nuclear stopping, the collective nuclear flow as well as the balance
 energy and pre-equilibrium nucleon emission. Bao An-Li et. al. have found
 that the mean nuclear field domains completely the density fluctuations as
 well as the isospin fluctuations,  meanwhile two-body scattering meaningfully
  influences in the subsequent isospin growth. Both fluctuations are increased
 in the dynamically instable systems meanwhile in the chemically instable 
systems only the density fluctuations are increased. The magnitude of both 
fluctuations diminishes when increases the isospin asymmetry resulting from the
 decrease of the isoscalar attractive mean field, due to the increase of the neutron
 symmetric repulsive potential in neutron rich nuclear matter \cite{4}. 
   Beaulieu et. al. analyzed experimental results of heavy ions collisions and
 found long relaxation times for surface fragment emissions and short relaxation
 times when fragments are emitted from the bulk  \cite{5}. In the soft region of the state 
equation, the pressure  change is much smaller and hence a small speed of sound  is
 attained since it depends directly on the pressure energy gradient: $V^2=\frac{dP}{de}$ . 

In this soft region, the collective dynamics of dense and hot matter formed in heavy
 ion collisions can be determinant. As a matter of fact, a small speed of sound delays 
the expansion of compressed matter and leads a lighter transverse collective flow \cite{6}. 
As shown via hydrodynamic models, heavy ion collisions reach the soft region of the
 state equation when the incident energy varies, building up a fireball that lasts a long
 time, leading to a minimum in the energy dependence of the collective transverse flow \cite{7}.  

\section{  Latino Model.}

   LATINO model \cite{8} uses semi-classical approximation to simulate 
intermediate energy heavy ions collisions and replicates binary interaction 
via Pandharipande potential. This potential is built up with a linear combination
 of Yukawa potentials, whose coefficients are fitted in order reproduce the
 properties of nuclear matter. Clusters are recognized using an Early Cluster
 Recognition Algorithm, that optimizes the configuration in the energy space. 
Ground states are produced generating a random configuration in phase space, 
gradually reducing the speed of the particles confined into a parabolic potential, 
until the theoretical binding energy is reached. Projectile kinetic energy varies
 in a range going from 800 MeV up to 2000 MeV in the center of mass system, 
with  200 replicas for each energy. Numerical integration of the equations of motion
 is performed with a Verlet algorithm with time intervals that ensure energy 
conservation up to 0.05
 converge to a power law in this range of intermediate energies \cite{9}.

\section{Methodology}

   In the Minimum Spanning Tree in Energy Algorithm (MSTE) , 
a given set of particles i,j,...,k belongs to the same cluster  $C_i$
whenever:
$ \forall i \in C_i,$   there exists $j \in C_i / e_{ij} \le 0$
where  $e_{ij}= V(r_{ij})+ ( \mathbf p_{i}- \mathbf p_{j})^2 / 4 \mu $
 , and $ \mu $ is the reduced mass of the pair {i,j}.

  Fragment formation time, $t_{ff}$, can be defined as the time when the system
 breaks up in a defined way, namely after $t_{ff}$  fragments evaporate only 
some monomers. In order to estimate $t_{ff}$  the "Microscopic Persistence
 Coefficient" $P$  is used, \cite{10}:
\begin{equation}  P \left[ X,Y \right] = 
{\frac {1}{ \sum_{cluster} n_{i}} }     \sum_{cluster} n_{i} \frac{a_{i}}{b_{i}}
\end{equation}  
where $ X \equiv \{ C_i \}$  and $Y \equiv \{ C'_i \}$  are two partitions, $b_i$ is the number of 
particle pairs belonging to cluster $C_i$ of partition X, $a_i$  is the number
 of particle pairs belonging to cluster $C_i$ and also belonging to a given
 cluster  $C'_j$  of partition Y,  $n_i$ is the number of particles in cluster $C_i$ .

   It is convenient to study the time evolution of the quantities:

\begin{equation} \hat P^{+} \left[ X(t) \right] \equiv
\left\langle P \left[ X(t), X(t \to \infty ) \right] 
\right\rangle_{collisions}
\end{equation}
\begin{equation}
 \hat P^{-} \left[ X(t) \right] \equiv 
\left\langle P \left[ X(t \to \infty ), X(t) \right] 
\right\rangle_{collisions}
\end{equation}
\begin{equation}
 \hat P^{dt} \left[ X(t) \right] \equiv
\left\langle P \left[ X(t), X(t+dt ) \right] 
\right\rangle_{collisions}
\end{equation}
where $X(t)$ is a partition at time t,  $X(t \to \infty)$ is an asymptotic partition, 
and   represents an average on all the collisions.  is a 
partition identical to the partition   , except for the fact that a nucleon
 has been evaporated in each cluster. 

\begin{equation}
 \lambda =\eval{ (n/p)}_{y > 0} / \eval{ (n/p) }_{y \le 0} 
\end{equation}
where $\eval{ (n/p)}_{y > 0}$ and $\eval{ (n/p) }_{y \le 0}$ 
are the ratios between the number of neutrons and protons moving forward and backwards, respectively.
   Dynamical instability is set when the squared speed of sound is negative. Adiabatic speed of sound is given by:
\begin{equation}
 V^2_C =(1/m)   \left[ ( \partial P/ \partial \rho)\rangle_{S} \right]
\end{equation}
\begin{equation}
 V^2_C =(1/m)   \left[ (10/9) <E_K> + a ( \rho / \rho_0 ) + b ( \rho / \rho_0 )^{\sigma}
 \right]
\end{equation}
where $<E_K> $ is the mean kinetic energy per nucleon, a=-358.1 MeV, b=304.8 MeV and $\sigma = 7/6$ 
are the parameters of the soft state equation. When $ V^2_C <0$ nuclear matter is instable
 with respect to density fluctuations, leading to dynamical instabilities \cite{11}.
   Heavy residuals thermal equilibrium can be examined studying the quadrupolar
 momentum $Q_{ZZ}$ defined by:
\begin{equation}
Q_{ZZ}  = \int dr dp   (2 \pi)^3 \left[ 2 p^2_z - p^2_y p^2_x \right]
 f(\overrightarrow r,\overrightarrow p,t)
\end{equation}
where $f( \overrightarrow r,\overrightarrow p,t)$  is Wigner function. Clearly $Q_{ZZ}=0$ is a necessary
 but not sufficient condition for thermal equilibrium \cite{12}. 
   Mean Velocity Transfer is given by:
\begin{equation}
MVT_i  = \sum_k \abs{ V_{ki} (t+ dt) - V_{ki} (t) }
\end{equation}

\section{Relaxation and Instabilities.}

   Collision Disordered Mode is the only one present in the uniformly excited systems, 
and is responsible of fragment production as well as the exterior flow dispersing the
 fragments \cite{13,14}. MSTE can be used to study biggest fragment size, total multiplicity
 and persistent coefficients. 

Figs.  1a, b and c show time evolution of these three quantities. Due to the initial
 violent collision, some particles attain lots of energy and are unable to be bound to 
a fragment even when these particles might be inside or close to a fragment. This 
reduces the size of the biggest MSTE cluster in this early stage and increases the
 total multiplicity.
    Fig, 2 shows Ni+Ni collision stages for an energy of 1400 MeV. The first stage
 ends up with a peak of kinetic energy transported by the promptly emitted particles, 
KE(PEP), signed by the peak of the wide continuous line. The second stage finishes
 up with the attenuation of the intermediate fragment production (IMF) , shown as a
 peak in the continuous curve. Between these two peaks, another peak is seen in 
the intermittent wide curve of the Mean Velocity Transverse (MVT), signing the
 onset of the instabilities. As can be seen, MVT produce instabilities leading to light
 particle emission and subsequent intermediate fragment emission (IMF). Once the
 continuous curve is attenuated, a final stage is installed with slow light fragment 
emission and system freezing-up. Persistent coefficients $P^+$  and $P^-$ computed 
using MSTE partitions,  can be sed to understand how the partition attains its
 microscopic composition at an asymptotic time.

   In this study dynamical, thermal and chemical instability signals are used to
 detect the relaxation times for several projectile energies, that are expected to
 reach a peak in the second collision stage and attenuate at the third collision stage.
   Fig. 3 shows the time evolution of the parameter $\lambda$ with a peak signing the
 entrance into the chemical instability zone, for the Ni+Ni central collision with 
a projectile energy of 1400 MeV. When the system attains chemical equilibrium, 
the parameter $\lambda$reaches a value close to zero. Fig. 4 shows the time evolution of
 the normalized quadrupole momentum with a peak signing the entrance into the
 thermal instability zone, for the central collision Ni+Ni, for a projectile energy of
 1400 MeV.  When the system attains thermal equilibrium, the normalized 
quadrupolar momentum reaches  a value close to zero. Fig. 5 shows time evolution
 of the squared adiabatic speed of sound with a peak signing the entrance into the
 dynamical instability zone, for Ni+Ni central collision with a projectile energy of 
1400 MeV. When the system reaches thermal equilibrium, the squared adiabatic 
sped of sound reaches a value close to zero.

\section{Conclusions.}

   Relaxation times obtained via these three signatures show that when the 
projectile energy is slow, the most important relaxation time is the thermal one, 
as expected since the available energy is spent in deformation and fragments are
 emitted from the surface (Fig. 6). But when the projectile energy is high, fragments
 are emitted from the bulk, and the three relaxation times are the same. Therefore, 
we expect to obtain low temperatures when the residual is big, since typically this
 case corresponds to low projectile energies and low excitations. Fig. 7 shows the
 time evolution for a typical experiment of the central collision Ni+Ni with a low
 projectile energy equal to 1100 MeV, 
where as can be seen, the compound suffers deformations up to an asymptotic state
 and a long quadrupolar relaxation time is expected. In a similar fashion, we expect
 high temperatures for a high excitation and a high projectile energy, since the
 relaxation times are short and the system disassembles almost immediately, spending
 the available energy on heating the system and leading to higher temperatures. When
 the projectile energy is increased, fragments are no longer emitted from the surface
 but from the bulk emission, and the dynamical behavior of the system is expected to
 change, providing different limit temperatures and distinct residual sizes. Work
 supported by the National Science Foundation (PHY-96-00038), Universidad de
 Buenos Aires (EX-070, Grant No. TW98, CONICET Grant No. PIP 4436/96), and
 Universidad Aut\'onoma Metropolitana-Azcapotzalco (Laboratorio de C\'omputo Intensivo).


\begin{thebibliography}{10}
 \bibitem{1}
 G.F. Bertsch and P.J. Siemens, {\it Phys. Lett}.B{\bf 126}, 9 (1983).
 \bibitem{2}
Bao-An Li {\it  et. al}., {\it Nucl.Phys.} A {\bf 630} 556 (1998). 
 \bibitem{3}
H. M\"uller {\it  et. al},{\it  Phys. Rev.}C{\bf 52}, 2072 (1995).
 \bibitem{4}
Bao-An Li {\it  et. al},{\it  Phys. Rev.} C. (2001).
 \bibitem{5}
Beaulieu {\it  et. al}.,{\it  Phys.Rev.Lett.} {\bf  84} 5971 (2000).
 \bibitem{6}
Bao-An Li {\it et. al}.,{\it  Phys. Rev.} C (1998).
 \bibitem{7}
C.M. Hung {\it et. al}, {\it Phys. Rev. Lett} {\bf .75}, 4003 (1995).
 \bibitem{8}
 A. Barra\~n\'on {\it et. al},{\it  Rev. Mex. F\'{\i}s.}{\bf 45}, 110 (1999).
 \bibitem{9}
A. Barra\~n\'on  {\it et. al.},  {\it Rev. Mex. F\'{\i}s}. {\bf 47}, 93 (2001). A. Barra\~n\'on {\it et. al.}.,  
{\it Heavy Ion Phys.} {\bf 17}-1, 59 (2003).
 \bibitem{10}
A. Strachan and C. O. Dorso, {\it Phys. Rev.} C{\bf 59}, 285(1999).
 \bibitem{11}
W. Bauer {\it et. al.}, {\it Phys. Rev. Lett.}C{\bf 69}, 1888 (1992).
 \bibitem{12}
Bao-An Li and S. Yenello,  {\it Phys.Rev.}C{\bf 52} 1746 (1995).
 \bibitem{13}
A. Strachan and C. O. Dorso, {\it Phys. Rev}.C{\bf 55}, 775(1997).
 \bibitem{14}
C. O. Dorso and J. Randrup, {\it Phys. Lett.}B{\bf 301}, 328(1993).

\end{thebibliography}
\end{document}